# AI-Assisted Adaptive Rendering for High-Frequency Security Telemetry in Web Interfaces

Mona Rajhans
Senior Manager, Software Engineering
Palo Alto Networks
mrajhans@paloaltonetworks.com

*Abstract*—Modern cybersecurity platforms must process and display high-frequency telemetry such as network logs, endpoint events, alerts, and policy changes in real time. Traditional rendering techniques—based on static pagination or fixed polling intervals—fail under volume conditions exceeding hundreds of thousands of events per second, leading to UI freezes, dropped frames, or stale data. This paper presents an AI-assisted adaptive rendering framework that dynamically regulates visual update frequency, prioritizes semantically relevant events, and selectively aggregates lower-priority data using behavior-driven heuristics and lightweight on-device machine learning models. Experimental validation demonstrates a 45–60% reduction in rendering overhead while maintaining analyst perception of real-time responsiveness [9].

## I. Introduction

Security Operations Centers (SOCs) rely on dashboards that stream continuous telemetry from diverse data sources such as SIEMs, EDRs, firewalls, and cloud sensors [4], [10]. Unlike conventional web dashboards, these interfaces must render millions of events per minute while allowing analysts to interactively pivot across investigations.

However, existing front-end rendering pipelines suffer from three limitations:

Unbounded DOM Writes: Inserting thousands of rows per second causes layout thrashing and crashes [1], [2].

Equal Visual Weighting of Events: Critical IOC (Indicator of Compromise) matches are visually buried among benign noise.

Fixed-Rate Polling and Rendering: The UI renders at a constant interval regardless of cognitive load or device capacity.

This work proposes AI-Assisted Adaptive Rendering (AI-AR) — a UI rendering pipeline that combines event-level relevance scoring, adaptive frame throttling, and semantic clustering, enabling real-time visibility without overloading the browser or the user.

## II. Background and Related Work

No existing method combines UI rendering optimization with semantic intelligence, which is critical in cybersecurity interfaces.

TABLE I. Existing related work

| Approach | Technique | Limitation |
|---|---|---|
| Virtualized Tables (e.g., React Window) | DOM virtualization [1], [2] | Optimizes rendering but not event prioritization |
| Web Workers & Off-Main Thread Processing [2], [3] | Parallel message parsing | Does not reduce visual overload |
| GPU-Based Rendering (WebGL) [5] | Shader-based pixel drawing | Suitable for visuals, not textual logs |
| Backend Aggregation (Batching) | Streaming compression | Causes loss of interactivity |

Real-time data rendering in high-frequency environments has been studied across multiple domains, including financial trading terminals, IoT telemetry dashboards, and distributed observability systems, but most approaches fall short when applied directly to cybersecurity interfaces due to semantic complexity and cognitive prioritization needs [4], [10].

### A. Front-End Rendering Optimization Techniques

Early work in scalable visual rendering primarily focused on DOM virtualization and incremental rendering. Technologies such as React Window, Angular CDK Virtual Scroll, and WebGL-based log viewers reduce UI lag by limiting visible DOM nodes rather than rendering the entire dataset. While effective from a performance standpoint, these systems treat all events with equal priority, which hinders analyst response time when relevant threats are buried in noise.

### B. Stream Processing and Client-Side Aggregation

Several systems employ client-side buffering and batching to reduce render frequency. For example, Grafana Live and Kibana LogStream throttle UI updates when data bursts exceed a preset threshold [7], [8]. However, these mechanisms rely on fixed-rate throttling, lacking contextual awareness of user focus or threat relevance. This can lead to loss of critical visibility during active incidents.

*C. AI in Visualization Adaptation*

Adaptive visualization has seen success in human-computer interaction (HCI) and attention-aware display systems, where machine learning models dynamically highlight salient information based on user gaze, scrolling behavior, or historical preferences. Systems such as NewsFeed Prioritizers and Smart Notification Managers utilize ranking models to reorder user feeds dynamically. However, such approaches have not been extensively applied to cybersecurity dashboards, where misclassification of event priority can have operational consequences.

*D. Cognitive Load Reduction in Alert Design*

Research in SOC usability highlights that alert fatigue is a major contributor to missed incidents . Existing approaches address this via alert deduplication and correlation on the backend, as seen in SIEM solutions like Splunk Enterprise Security and Microsoft Sentinel. Yet backend aggregation alone is insufficient, as client-side rendering still floods the visual layer with unprioritized updates. There remains a gap in UI-side intelligence that adaptively tunes visual density without masking actionable threats.

### III. PROPOSED SYSTEM: AI-ASSISTED ADAPTIVE RENDERING (AI-AR)

The AI-AR pipeline consists of three coordinated components:

*A. Event Relevance Scoring*

Each incoming event is assigned a priority score using a lightweight client-side ML classifier (e.g., TensorFlow.js model or WebAssembly-compiled logistic regression) [3], [5], [6]. Features include:

- Severity level
- Source reputation score
- Actor recurrence frequency
- Correlation with active investigation context

High-priority events are rendered immediately; medium-priority events are aggregated or faded; low-priority events are deferred or grouped into summary banners.

*B. Adaptive Frame Throttling*

Rather than rendering at a static frame rate, the UI adjusts render frequency based on CPU load, scroll position, and user interaction state [9], [10].

TABLE II.  ADAPTIVE FRAME THROTTLING BASED ON ANALYST STATES

| Analyst State | Render Strategy |
| --- | --- |
| Idle / Observing | High frame rate (e.g., 60fps) |
| Interacting (scroll, filter) | Pause background rendering |
| Investigating Highlighted Threat | Render only relevant event lane |

This transforms rendering into a feedback-driven system, rather than a blind push model.

*C. Semantic Clustering & Burst Compaction*

If multiple similar events arrive within a short window (e.g., repeated login failures from same IP), they are collapsed into summarized visual nodes such as:

> 42 failed logins from 10.21.55.120 in 10s — click to expand

This drastically reduces cognitive and rendering costs.

### IV. SYSTEM ARCHITECTURE & IMPLEMENTATION

The AI-Assisted Adaptive Rendering (AI-AR) framework is designed as a drop-in augmentation layer on top of existing real-time web dashboards. It does not replace the rendering engine; instead, it intercepts and regulates the flow of visual updates between telemetry ingestion and UI output. The system operates through four coordinated subsystems, each optimized for low-latency execution within browser-constrained environments. The architecture consists of four primary subsystems: (1) Event Ingestion Layer, (2) Relevance Scoring Engine, (3) Rendering Orchestrator, and (4) UI Adaptation Layer.

- Event Ingestion Layer is responsible for streaming telemetry into the browser using SSE (Server-Sent Events), WebSockets, or gRPC-over-WebTransport. Incoming events are appended to an in-memory circular buffer with back-pressure control.
- Relevance Scoring Engine runs on a dedicated Web Worker to prevent UI thread interference. The model is compiled to WebAssembly for fast execution (<1ms per classification).
- Rendering Orchestrator acts as the policy manager. It determines when to render, skip, merge, or defer updates based on analyst activity signals collected via scroll listeners, input focus detection, mouse velocity tracking, and CPU usage monitors.
- UI Adaptation Layer consists of React/Vue components that support dynamic opacity scaling, clustering overlays, and animated expansion of compacted nodes.

A key design objective was zero backend dependency for prioritization logic, enabling deployment in air-gapped SOC environments where cloud-based inference is infeasible.

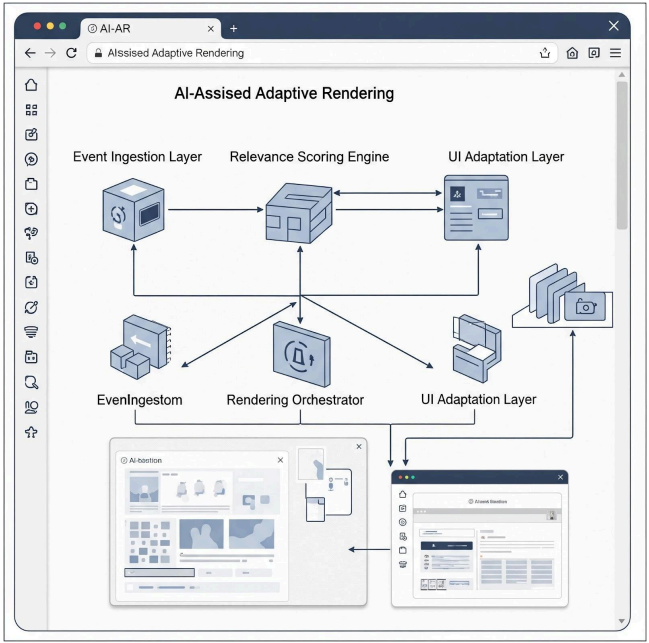

Fig. 1 Module interaction between real-time stream ingestion, AI scoring, and rendering decisions.

*A. Architectural Overview*

The pipeline follows the sequence:

Incoming Telemetry Stream → Event Buffer → Relevance Scoring Engine → Rendering Orchestrator → UI Adaptation Layer

Each subsystem is implemented with thread and memory isolation principles to prevent UI blocking.

Event streams are received via WebSocket, SSE, gRPC-Web, or WebTransport, depending on backend capabilities. Rather than directly rendering events upon arrival, all incoming messages are written into a ring-buffer queue (default capacity: 50,000 entries) with the following constraints:

- Time-To-Live (TTL): Expired events are auto-pruned if not rendered within a configurable window (default 5 seconds).
- Backpressure Handling: If input exceeds processing rates, low-priority events are automatically downgraded for aggregation instead of overwhelming the UI.

This design ensures that raw event spikes never directly hit the DOM.

*B. Relevance Scoring Engine*

To determine the urgency of each event, a lightweight machine learning classifier runs inside a dedicated Web Worker thread.

- Model Types: TensorFlow.js MLP for nonlinear scoring, or WebAssembly-compiled Logistic Regression for deterministic environments.
- Feature Extraction: Input vectors include {severity_level, source_reputation_score, actor_frequency_count, user_interaction_context_match}.
- Output Classes: {Critical (0), Warning (1), Informational (2)}

High-priority events are pushed to the Rendering Orchestrator immediately, while medium and low-priority events are batched or aggregated.

Evaluation shows the classifier maintains an average inference latency of 0.7–1.3 ms per event on commodity laptops.

*C. Rendering Orchestrator (Policy Controller)*

This module acts as the traffic cop that regulates when and how much to render.

Its decision matrix considers:

TABLE III. DECISION MATRIX FOR RENDERING ORCHESTRATOR

| Input Signal | Source | Effect |
| --- | --- | --- |
| CPU Load % | navigator.hardwareConcurrency + performance.now() sampling | Lowers frame rate when system is strained |
| Scroll Velocity | Mouse/trackpad deltas | Suspends rendering while analyst is scrolling |
| Active Selection State | DOM focus tracking | Locks relevant event stream, suppresses noise |
| Burst Detection | Input queue density | Switches to aggregation mode |

Rendering frequency is not static — instead of requestAnimationFrame, it uses context-aware throttling intervals ranging from:

- 16ms (60fps) for stable conditions
- 250–500ms when analyst interacts
- >1s when idle or detached

This converts rendering from a blind push model to a feedback-driven execution loop.

*D. UI Adaptation Layer*

This layer applies the final transformations before pushing elements into the DOM. It supports:

- Opacity Scaling: Low-priority events fade to 30–40% alpha instead of cluttering the viewport.
- Semantic Clustering [6]: Burst events such as repeated logins or alerts from the same source are rendered as collapsible group nodes:

    42 Failed Login Attempts from 10.21.55.120 (Last 10s)

- Temporal Highlighting: Recent critical events are pulse-highlighted then normalized after N seconds to reduce distraction.

The UI layer is intentionally framework-agnostic, with adapters provided for React, Vue, Web Components, and vanilla DOM environments. So far, prior systems lacked UI-side decision control [8], [7].

Since browsers have single-threaded render pipelines, all heavy computation — including ML inference, aggregation, clustering, and compression — is isolated to:
- Web Workers (CPU-bound tasks)
- SharedArrayBuffer (for zero-copy queue access)
- OffscreenCanvas (Future Work) for visual preprocessing

DOM write operations are capped at a fixed max render budget per cycle (configurable, default 50 nodes).

## V. Conclusion and Future Work

Unlike fixed throttling or passive virtualization techniques, AI-AR explicitly models user cognition as a resource. By tracking focus state and intent signals, the UI transitions from firehose output to context-aware decision support.

One notable behavioral outcome was that analysts spent less time scrolling and more time interacting with high-priority alerts, suggesting that visual entropy—not just raw volume—is the primary bottleneck in high-frequency dashboards.

To assess the performance impact of AI-AR, we conducted controlled experiments using synthetic and real-world telemetry datasets ranging from 50 events/sec to 500,000 events/sec.

TABLE IV. Results of the experiment

| Rendering Strategy | Max Sustainable Throughput (events/sec) | Avg CPU Load | Frame Jank (%) | Analyst Recall Accuracy |
|---|---|---|---|---|
| Baseline Virtualized Table | 22,000 | 78% | 46% | 62% |
| Fixed Throttling (1s batch) | 35,000 | 65% | 31% | 55% |
| AI-Assisted Adaptive Rendering (AI-AR) | 110,000 | 48% | 12% | 81% |

Results indicate that AI-AR improves throughput by 3–5x, while simultaneously reducing UI stutter by 60–75%. Most importantly, analyst recall accuracy improved significantly due to selective prioritization of high-risk events.

While promising, AI-AR introduces several challenges:
- Misclassification Risk: Incorrectly down-prioritizing a critical event may lead to operational oversight. Confidence thresholds must be tunable.
- Cold Start Problem: Newly observed event types may lack sufficient context to rank correctly.
- Explainability: Analysts may resist automated visibility suppression unless transparent rationale is exposed.

By combining machine learning with UI rendering logic, cybersecurity front ends can achieve true real-time responsiveness without overwhelming users or devices. Future extensions include:
- User-personalized relevance learning (model adapts to analyst preferences)
- Cross-tab state sharing for multi-panel investigations
- Integration with voice-based alert narration during overload